\documentstyle[12pt]{article}

\newcommand{\be}{\begin{equation}}
\newcommand{\ee}{\end{equation}}
\newcommand{\ba}{\begin{eqnarray}}
\newcommand{\ea}{\end{eqnarray}}

\input{tcilatex}
\begin{document}

\begin{titlepage}
\pagestyle{empty}
\vspace{1.0in}
\begin{flushright}
hep-th/9707247   \\
SNUTP/97-106 \\
July 1997
\end{flushright}
\vspace{.1in}
\begin{center}
\begin{Large}
{\bf Superpotentials of 
$N=1$ Supersymmetric } \\
{\bf Gauge Theories from M-theory}  
\end{Large}
\vskip 0.5in
Soonkeon Nam$^{\S }$,
Kyungho Oh$^{* }$,
Sang-Jin Sin$^{\dagger}$ 
\vskip 0.2in  
\begin{center} {{\it $^\S$ Dept. of Physics, Kyung Hee University,
Seoul 130-701, Korea \\ 
{\tt nam@nms.kyunghee.ac.kr}\\}} 
{{\it $^*$ Dept. of Mathematics, 
University of Missouri-St. Louis, \\  
\vskip -0.2cm
{{\it St. Louis, Missouri 63121, USA}}\\  
{\tt oh@arch.umsl.edu }\\}}
{{\it $^\dagger$ Dept. of Physics, Hanyang University,
Seoul 133-791, Korea\\
{\tt sjs@dirac.hanyang.ac.kr}}}
\end{center}
\end{center}
\vspace{0.5 cm}

\begin{abstract}
We consider brane configurations in M-theory describing $N=1$ 
supersymmetric  gauge theories and 
using the parametric representation of the brane configurations,
we calculate the  superpotentials for various cases
including multiple gauge groups or fermions.
For  $SU(n)$ $N=1$ SQCD with $N_f$ fermion case ($N_f < N_c)$,
we find that the superpotential from M-theory
and the gauge theory agree precisely. 
This gives a direct evidence of the validity of Witten's M-theory 
method for calculating the superpotential.
\end{abstract}
\end{titlepage}

\section{Introduction}

The idea of D-brane\cite{pol} opened up a new and surprisingly simple way to
communicate between super symmetric (SUSY) gauge theories and the
superstring theories. Many intriguing results about the field theories have
been achieved by investigating gauge theories formulated on the world-volume
of branes in string theories[2-11] and M-theory[12-16]. In a recent paper
\cite{witten1}, Witten provided solutions of $N=2$ SUSY gauge theories in
four dimension\cite{sw1,sw2} by reinterpreting configurations of fourbranes
and fivebranes in type II superstring as the branes in M theory. Moreover,
in a subsequent paper\cite{witten2}, he showed how some of the outstanding
problems in particle physics such as quark confinement and chiral symmetry
breaking, can be approached from the M-theory point of view. There, he also
suggested a way to calculate the superpotential directly from a brane
configuration, thereby he gave a direct evaluation of the tension of the
domain wall\cite{cs}. Though very insightful, no direct comparison with the
gauge theory result was given.

One of the main goal of this paper is to give a direct evidence for the
validity of Witten's method by extending his result to the cases where gauge
theory results are available. We will show that the minimum of the
superpotential of the $N=1$ $SU(n)$ gauge theory superpotential agrees with
the value evaluated with the M-theory method. We will also generalize and
obtain the superpotentials for more general situations with product gauge
groups, which corresponds to the brane configurations where there are $n+1$
fivebranes. To motivate the study of multiple branes we recall the works of
Elitzur et al. who have considered $k$ `coincident' NS fivebranes linked to
an `coincident' NS fivebrane linked by $N_c$ D fourbranes\cite{egkrs}.

\section{The Brane Configuration}

To fix the notation, we begin with brane configurations in type IIA
superstring theory. Let the spacetime coordinates be ($x^0, x^1, \ldots ,
x^9 $) on ${}^{10}$. Our basic brane configuration contains three kinds of
branes: NS fivebranes whose world volume occupies ($x^0, x^1, x^2, x^3, x^4,
x^5$) spacetime coordinates, D fourbranes occupying $(x^0, \cdots, x^3, x^6 )
$, and D sixbranes occupying $(x^0, \cdots, x^3, x^7, x^8, x^9 )$.

The basic brane configuration consists of $N_c$ fourbranes suspended between
two parallel fivebranes in Type IIA superstring theory which gives a
representation of $N=2$ SUSY QCD (SQCD) in four dimensions $(x^0, \cdots,
x^3)$ with gauge group $SU(N_c)$. Also $N_f$ D sixbranes could be added and
this gives $N_f$ hypermultiplets in the fundamental representation of the
gauge group $SU(N_c)$.

One of the limitation of type IIA superstring theory is that a fourbrane
ending on a fivebrane defies any explicit conformal field theory
description. However, by going to M theory, the picture becomes more
unified. A type IIA fourbrane becomes an M theory fivebrane that is wrapped
over the $11^{{\rm th}}$ dimension $^1$. Thus D fourbranes and NS fivebranes
come from the same basic object in M theory. This $N=2$ SUSY description can
be broken to $N=1$ by turning on a mass for the adjoint chiral superfield in
the $N=2$ vector multiplet.

Let us now review some of the Witten's construction\cite{witten1,witten2} of
the parametric representation of the brane configuration in M theory for $
N=1 $ SQCD. To set up the brane configurations in complex geometry, we
introduce complex coordinates $v=x^4+ix^5$, $w=x^7+ix^8$, $t=\exp
(-(x^6+ix^{10})/R)$. The brane configuration for $SU(n)$ gauge theory is
given by the Seiberg-Witten curve\cite{sw1} 
\begin{equation}
{t^2+P_n(v)t+1 = 0},
\end{equation}
in $v-t$ space when $w=0$. Here $P_n(v)$ is a polynomial of the form 
\begin{equation}
{P_n(v)=v^n +u_2v^{n-2}+\dots +u_n,}
\end{equation}
where the $u_i$'s are the ``order parameters'' of the theory.

The brane configuration associated to $N=1$ SUSY can be obtained by
``rotating'' the brane configuration that describes $SU(n)$ gauge theory
with $N=2$ SUSY, in the sense of Ref.\cite{rotat}. The rotation is possible
only when all the 1-cycles on the corresponding curve are degenerate. This
is why the $N=1$ curve is rational for the two NS fivebrane case\cite
{witten2}. However for more than two NS fivebranes, we can ask whether the
curve is generically rational. By reversing the argument we can say that $
N=2 $ SUSY forbids the rotation of the NS fivebranes. Once $N=2$ SUSY is
broken, however, there is nothing that freezes the rotating degrees of
freedom. This means that in generic configuration the NS fivebranes are
rotated to one another by certain angles. Therefore the curves with positive
genus correspond very special non-generic cases. Hence the generic M-brane
configuration corresponds to a curve $\Sigma$ of genus zero, which has
parametric representations. Namely, $\Sigma$ can be identified by a
holomorphic mapping with a Riemann sphere with some points deleted. If $
\lambda$ is the complex coordinate of $\Sigma$, the holomorphic functions $
v(\lambda),w(\lambda),$ and $t(\lambda)$ on $\Sigma$ are given by rational
functions of $\lambda$.

Since $v$ has only poles at ends of fivebranes, we can put $
v=\lambda+c\lambda^{-1}$, for some constant $c$. Also, $t$ can go to zero or
infinity only at poles of $v$, so $t$ is a constant multiple of a power of $
\lambda$, i.e. $t\sim \lambda^n$. We also get $c^n=1$ from eq.(1), and for
each choice of $c$ the polynomial $P_n$ is uniquely determined. After the
rotation, $w$, instead of being zero, should be a non-zero holomorphic
function on $\Sigma$. Rotating only one fivebrane, $w$ should get a pole
only at one end of $\Sigma$ and should vanish at the other end. So we can
set $w=\zeta \lambda^{-1}$ for some complex constant $\zeta$. As $\lambda$
goes to zero $w/v \to \zeta/c $, therefore it can be regarded as a tangent
of the rotation angle, namely 
\begin{equation}
\tan \theta=|\zeta/c| .
\end{equation}
For the rotation of $\theta=\pi/2$, one should set $c=0$, because $v$ and $w$
do not have a common pole. Therefore the parametric equation for the curve
becomes; $v=\lambda$, $w=\zeta \lambda^{-1}$, $t=\lambda^n$.

Now we give a parametric representation of the brane configuration with more
than two NS fivebranes. Suppose NS fivebranes are located at $\lambda =
\lambda_\alpha$, $\alpha=0,1,\cdots,n$ at arbitrary angles in $v-w$ space,
with $k_\alpha$ D fourbranes suspended between $\alpha -1 ^{{\rm th}}$ and $
\alpha^{{\rm th}}$ fivebranes. Thus the functions $v$ and $w$ on $\Sigma$
have poles at $\lambda_\alpha$, $\alpha=0,1,\cdots,n$. The function $t$ on $
\Sigma$ will have a zero or pole according to the direction the NS fivebrane
is bent. If there are more D fourbranes in the left than right it will bend
to the right, therefore it will have a zero and vice versa. In fact $t\sim
v^{a_\alpha}$ as $v \to \infty$ where $a_{\alpha} = k_{\alpha+1}-k_{\alpha}$
. Putting all these together, we have 
\begin{eqnarray}
&&v=\sum_{\alpha=0}^{n} \frac{c_{\alpha}}{\lambda -\lambda_{\alpha}}, 
\nonumber \\
&&w=\sum_{\alpha=0}^{n} \frac{s_{\alpha}}{\lambda -\lambda_{\alpha}}, 
\nonumber \\
&&t=\prod_{\alpha=0}^{n}(\lambda -\lambda_{\alpha})^{-a_{\alpha}},
\end{eqnarray}
for some complex numbers $c_\alpha,\ s_\alpha$. We normalize the system by
setting $c_{0}=1$, $s_{0}=0$, so that $\lambda_0$ is the location of the
unrotated NS fivebrane. Then the rotation angles $\theta_{\alpha}$'s are
again given by 
\begin{equation}
\tan \theta_{\alpha}=\left| {\ {\frac{s_{\alpha}}{c_{\alpha}}}} \right|.
\end{equation}
Notice that Witten's choice corresponds to the special case where two poles
are chosen as $\lambda=0$ and $\infty$ and with rotation angle $\pi/2$.
Since it can be related to the $n=1$ case of above with 
$(\lambda_0,\lambda_1)=(0,1)$ 
by the M\"obius transformation 
$\lambda\to {\tilde \lambda}={\frac{\lambda-1}{\lambda}}$ 
followed by constant shifts
in $v$ and $w$, they are physically equivalent.

When one approaches the ends of the fivebranes, $v$ and $w$ go to infinity.
Since the effect of the rotation can be ignored, the equation for the $N=2$
Seiberg-Witten curve 
\[
t^{n+1}+P_{k_{1}}(v)t^{n}+\cdots +1=0, 
\]
must be satisfied near the ends. This determines some of the moduli of the $
N=1$ curve in terms of the moduli of the polynomial equation for $N=2$ case.
As $\lambda $ approaches $\lambda _{\alpha }$ the parametrization of the
curve is asymptotically given by, 
\begin{eqnarray}
&&v\to \frac{c_{\alpha }}{(\lambda -\lambda _{\alpha })},  \nonumber \\
&&w\to \frac{s_{\alpha }}{(\lambda -\lambda _{\alpha })},  \nonumber \\
&&t\to (\lambda -\lambda _{\alpha })^{-a_{\alpha }}\prod_{\beta \ne \alpha
}(\lambda _{\alpha }-\lambda _{\beta })^{-a_{\beta }}.
\end{eqnarray}
Inserting these to eq.(6), we get the following: 
\[
c_{\alpha }^{a_{\alpha }}=-{\frac{p_{\alpha +1}}{p_{\alpha }}}\prod_{\beta
\ne \alpha }(\lambda _{\alpha }-\lambda _{\beta })^{-a_{\beta }},\ \ \
(\alpha =0,\cdots ,n), 
\]
where $p_{\alpha }$ denotes the coefficient of the leading power in $
P_{k_{\alpha }}(v)=p_{\alpha }v^{k_{\alpha }}+\cdots $. This determines the
coefficients of the poles of $v$ as function of the location of the
fivebranes and the information of $N=2$ theory up to the phase $e^{2\pi
i/(k_{\alpha +1}-k_{\alpha })}$. As Witten argued, $p_{\alpha }$'s are
parameters of the Lagrangian rather than a modulus, so we conclude that only
the locations of the NS fivebranes $\lambda _{\alpha }$'s and the parameter
for angles of the $\alpha +1^{{\rm th}}$ NS brane relative to the first one, 
$s_{\alpha }$'s, are the moduli of the $N=1$ theory.

Let us now discuss the symmetry of the curve. Although there is no remaining
symmetry for interacting branes, asymptotically, when $\lambda $ is close to 
$\lambda _{\alpha }$, we have following symmetry: 
\[
\prod_{\alpha =1}^{n}{\mathbf Z}_{|a_{\alpha }|}=
{\mathbf Z}_{k_{2}-k_{1}}\times \cdots \times 
{\mathbf Z}_{k_{n-1}-k_{n}}\times {\mathbf Z}_{k_{n}}.
\]
This corresponds to the rotational symmetries of the branes when $v$ goes to
infinity broken to the discrete symmetries due to the quantum effects. We
remark that this is the analogue of chiral symmetry for $SU(n)$ case in
which the chiral symmetry is ${\mathbf Z}_{n}$. The structure of the symmetry
manifestly reveal the consequence of the interactions between the branes.
The field theory analogues of this effect for the product gauge groups are
not fully studied. One technical remark is that in considering the symmetry
properties, it is convenient to locate the zeroth brane at the infinity so
that it looks special relative to the others.

\section{The Superpotential}

Given the brane configurations in M-theory, we calculate the superpotential
of the corresponding gauge theory following Witten's idea\cite{witten2}.
Consider in general M-theory compactification on ${\bf R}^4 \times X \times 
{\bf R}$ where $X$ is a Calabi-Yau threefold. Suppose in spacetime there are
fivebranes of the form ${\bf R}^4 \times \Sigma$, $\Sigma$ being a
two-dimensional real surface in $X$. Choose $\Sigma_0$ in the homology class
of $\Sigma$ in $H^2(X, {\bf R})$. Then there exists a three manifold $B$ and
a map $\Phi_B: B\to X$ such that the boundary of $B$ maps $\Sigma - \Sigma_0$
in $X$. Let $\Omega$ be the holomorphic three form on $X$. Then as Witten
suggested, the superpotential is given by 
\begin{eqnarray}
W(\Sigma) - W(\Sigma_0) = \int_B \Phi_B^* (\Omega).
\end{eqnarray}
This defines $W(\Sigma)$ up to an additive constant. The ambiguity in $
W(\Sigma)$ comes from the choices of $B$ and $\Sigma_0$. In general, the
periods of $\Omega$ also contribute to the ambiguity, but this
indeterminancy disappears since $H_3(X, {}) =0$ in our applications. The
condition $H_3(X, {}) =0$ also forces that the space of all possible $\Sigma$
is simply-connected. Thus a different choice of $B$ does not create any
additional constant.

To discuss the issue of the $\Sigma_0$ dependence, we consider two NS
fivebranes connected by $n$ D fourbranes and assume that two are at a
relative angle $\theta$. We take $X$ to be a flat Calabi-Yau manifold $Y$
with coordinates $v,w,$ and $t$. The holomorphic three form $\Omega$ on $Y$
is given by 
\begin{equation}
\Omega = R \frac{dt}{t}\wedge dv \wedge dw .
\end{equation}
It is chosen such that $\Omega \wedge \overline{\Omega}$ is the Riemannian
volume form for $Y$. Since we can choose the position of a fivebrane at $
\lambda=0$ without loss of generality, the configuration is parametrized by 
\begin{eqnarray}
&&v(\lambda)= \frac{1}{\lambda -\lambda_1}+\frac{c}{\lambda},  \nonumber \\
&&w(\lambda)= \frac{\zeta}{\lambda},  \nonumber \\
&&t(\lambda)= \left(\frac{\lambda}{\lambda -\lambda_1}\right)^n.
\end{eqnarray}

In fact the dependence of the superpotential on the choice of the $\Sigma_0$
can be a subtle issue. In principle, fixing a $\Sigma_0$ just amounts to
fixing the zero point of the superpotential. In practice, however, for a
given $\Sigma$ one has to choose an appropriate $\Sigma_0$ to expedite
calculation. Therefore we want to put $W(\Sigma_0)=0$ for a certain class of
surfaces. In the next paragraph, for the fixed $\Sigma$ given by eq.(12), we
choose ${\tilde\Sigma}_0$ and $\Sigma_0$ which look completely different,
and show by explicit calculation that they in fact give the same
superpotential.

To construct ${\tilde \Sigma}_0$ we first introduce a new variable 
${\tilde\lambda}={
\frac{\lambda}{\lambda-\lambda_{1}}}$ and take $\Sigma_\lambda$ to be the
complex $\lambda$-plane with $0$ and $\lambda_1$ deleted. 
We write ${\tilde\lambda}=\exp({\tilde\rho}+i{\tilde\theta})$, with ${\
\tilde\rho}$ and ${\tilde\theta}$ real, and pick an arbitrary smooth
function ${\tilde f}$ of a real variable such that ${\tilde f}({\tilde\rho}
)=1$ for ${\tilde\rho}>2$ and ${\tilde f}({\tilde\rho})=0$ for ${\tilde\rho}
<1$. Then we define the map $\Phi_0:\Sigma_\lambda \to Y$ by 
\begin{eqnarray}
&& v = {\frac{1}{\lambda-\lambda_1}}{\tilde f}({\tilde\rho}) +{\frac{c}{
\lambda}} {\tilde f}(-{\tilde\rho}),  \nonumber \\
&& w = {\tilde f}(-{\tilde\rho}) \frac{\zeta}{\lambda},  \nonumber \\
&& t = \left(\frac{\lambda}{\lambda -\lambda_1}\right)^n .
\end{eqnarray}
By construction ${\tilde \Sigma}_0$ is asymptotic at infinity to $\Sigma$. We now
introduce smooth bounded functions $g_{\pm}=g_{\pm}({\tilde\rho},\sigma)$
given by $g_{+}({\tilde\rho},1)=1$, $g_{+}({\tilde\rho},0)=f({\tilde\rho})$,
and $g_{+}({\tilde\rho},\sigma)=1$, and $g_{-}({\tilde\rho},\sigma)=g_{+}(-{
\tilde\rho},\sigma)$. The map $\Phi_B:B\to Y$ can then be defined by 
\begin{eqnarray}
&&v = {\frac{1}{\lambda-\lambda_1}} g_{+}({\tilde\rho},\sigma) + {\frac{c}{
\lambda}}g_-({\tilde\rho},\sigma),  \nonumber \\
&&w = {\frac{\zeta}{\lambda}} g_{-}({\tilde\rho},\sigma),  \nonumber \\
&&t = {\tilde\lambda}^{n}.
\end{eqnarray}
The superpotential now becomes 
\begin{equation}
W(\Sigma)-W({\tilde \Sigma}_0)=Rn\int_B {\frac{d{\tilde\lambda}}{{\tilde\lambda}}}\wedge
dv\wedge dw,
\end{equation}
and thus 
\begin{equation}
W(\Sigma)-W({\tilde \Sigma}_0)=iRn \int_0^1d\sigma\int_0^{2\pi} d{\tilde\theta}
\int_{-\infty}^\infty d{\tilde\rho}{\frac{\zeta}{\lambda_1^2}}{\frac{({\
\tilde\lambda}-1)^2 }{{\tilde\lambda}}} \left({\frac{\partial g_+}{
\partial\sigma}}{\frac{\partial g_-}{\partial\rho}} -{\frac{\partial g_+}{
\partial \rho}}{\frac{\partial g_-}{\partial\sigma}}\right).
\end{equation}
Notice that the $g_{\pm}$ is independent of ${\tilde\theta}$. Therefore the
integrand is sum of the terms whose integrals split into 
$\int d{\tilde\theta}$ and the rests. The final result is 
\begin{equation}
W(\Sigma)=-{\frac{4\pi i Rn\zeta }{\lambda_1^2}}.
\end{equation}

Now we choose a different $\Sigma_0$ given by the $\Phi_0$ which is
determined by continuous and smooth functions $f_0,f_1$. Let $
\lambda/\lambda_1=\exp(\rho+i\theta)$ and construct $f_i$'s such that $
f_0(\rho)=1$ if $\rho<-3$, $f_0=0$ if $\rho>-2$ and $f_1(\rho)=1$ if $
|\rho-1|<1 $, $f_1=0$ if $|\rho-1|>2 $. Construct the map $\Phi_0:
\Sigma_\lambda \to Y$ just as above by replacing ${\tilde f}(\rho)$ and 
${\tilde f}(-\rho)$ by $f_i(\rho)$'s. Then construct $g_i$, $i=0,1$ such that
they interpolate $f_i$ and constant function 1 continuously and smoothly.
Then we choose the map $\Phi_B: B\to Y$ as above by replacing $g_\pm$ by $
g_i $'s. Then the superpotential now becomes 
\begin{equation}
W(\Sigma)-W(\Sigma_0)=iRn\int \left[ \frac{1}{\lambda^2(\lambda-\lambda_1)}
- \frac{1}{\lambda(\lambda-\lambda_1)^2}\right] d\lambda\wedge dg_1 \wedge
dg_0 .
\end{equation}
This integral also can be evaluated easily by noticing that $g_i$'s do not
depend on the angular variable $\theta$, and is exactly the same as the
previous case. Although the bases of $f$'s and $\tilde f$'s, defined as the
regions in $\Sigma_\lambda$ where $\Sigma_0$ and ${\tilde \Sigma}_0$ 
respectively is
equal to $\Sigma$, look very different, $\Sigma_0$ and $\tilde \Sigma_0$ has
a common property. They both asymptotically approach to $\Sigma$ and have
necks that have zero thickness which makes the ${}_n$ invariance more
manifest. The general construction described below will respect this feature.

Notice that the superpotential is independent of the rotation angle for the
two fivebrane cases. There are overall factor 2 difference between the
previous result with that for Witten's configuration, which gives $2\pi$
rather than $4\pi$. This can be understood by noticing that in Witten's
choice, the pole at the infinity is truncated from the expression of $t$ so
that the contribution from that pole is not included.

Now let us calculate the superpotential for the general multi-brane cases,
given by eq.(4). We assume that $|\lambda_0| >|\lambda_1| >\cdots>|\lambda_n|$.
Let $\rho=\log |\lambda|$, $\rho_\alpha=\log |\lambda_\alpha|$ for $\alpha = 0,
\cdots, n$, and $\epsilon=\min_{\alpha=0, \dots,n }(
\rho_\alpha-\rho_{\alpha+1} ).$ We construct $\Sigma_0$ in terms of 
the functions $f_\alpha(\rho)$ which are defined by 
\begin{eqnarray}
&&f_\alpha(\rho) = 1 \quad {\rm if} \quad |\rho-\rho_\alpha|< {\frac{1}{3}}
\epsilon,  \nonumber \\
&&\ \ \qquad = 0 \quad {\rm if} \quad |\rho-\rho_\alpha|> {\frac{2}{3}}
\epsilon,
\end{eqnarray}
and $f_\alpha$ interpolate 0 and 1 in the region ${\frac{1}{3}}\epsilon <
|\rho-\rho_\alpha| < {\frac{2}{3}}\epsilon $. We define the homotopy
functions $g_\alpha(\rho,\sigma)$'s which interpolate $f_\alpha(\rho)$ and 1
such that $g_\alpha(\rho,0) = f_\alpha(\rho),$ $\ g_\alpha(\rho,1) = 1$. In
other words, each $f_\alpha$ has a circular strip containing the circle
passing the $\lambda_\alpha$ as its territory where it is 1 and rapidly dies
outside and no territories are overlapping so that $f_\alpha f_\beta=0$ for
any pair. Furthermore for later purpose we construct the $g_\alpha$'s such
that 
\begin{eqnarray}
g_\alpha(\rho_\gamma,\sigma)=g_\beta(\rho_\gamma,\sigma) :=h_\gamma (\sigma),
\end{eqnarray}
for any triple $\alpha, \beta, \gamma$ which are different to one another.
Now take a three manifold $B$ as the product of the $\lambda$ plane with all
the $\alpha$'s deleted and the $\sigma$ interval $0\leq \sigma \leq 1$. Then
define a map $\Phi:B\to Y$ by 
\begin{eqnarray}
&&v=\sum_{\alpha=0}^{n} \frac{c_{\alpha}}{\lambda -\lambda_{\alpha}}
g_{\alpha}(\rho,\sigma),  \nonumber \\
&&w=\sum_{\alpha=0}^{n} \frac{s_{\alpha}}{\lambda -\lambda_{\alpha}}
g_{\alpha}(\rho,\sigma),  \nonumber \\
&&t=\prod_{\alpha=0}^{n}(\lambda -\lambda_{\alpha})^{-a_{\alpha}}.
\end{eqnarray}

Now the superpotential can be written down as 
\begin{eqnarray}
&&W=W(\Sigma)-W(\Sigma_0)=R \int {\frac{dt}{t}}\wedge dv \wedge dw  \nonumber
\\
&& \ \ \ \ =R\sum_{\alpha,\beta,\gamma} c_\alpha s_\beta a_\gamma \int
dg_\alpha\wedge dg_\beta \wedge d\lambda I_{\alpha,\beta,\gamma}(\lambda),
\end{eqnarray}
where 
\[
I_{\alpha,\beta,\gamma}(\lambda) = \frac{1}{(\lambda
-\lambda_{\alpha})(\lambda -\lambda_{\beta}) (\lambda -\lambda_{\gamma})}. 
\]
Let $\lambda=|\lambda|e^{i\theta}$ and $z=e^{i\theta}$. Then we can first
evaluate the $d\lambda$ integral as $d\theta$ integral, which in turn can be
evaluated as contour integral on $z$ plane along the unit circle. Therefore 
\begin{eqnarray}
\int d \lambda I_{\alpha,\beta,\gamma}(\lambda) = \int |\lambda|dz \frac{1}{
(|\lambda|z -\lambda_{\alpha})(|\lambda|z -\lambda_{\beta}) (|\lambda|z
-\lambda_{\gamma})}.
\end{eqnarray}
Due to the presence of the homotopy factors, one can avoid $
\lambda_\alpha=\lambda_\beta$ However, we can have 
$\lambda_\gamma=\lambda_\alpha$ or $\lambda_\gamma=\lambda_\beta$.

\begin{itemize}
\item  The case with $\alpha \ne \gamma \ne \beta $
\end{itemize}

\begin{eqnarray}
\int d\lambda {I_{\alpha,\beta,\gamma}(\lambda) }=2\pi i \big[ 
\Lambda_\alpha\Theta(\rho-\rho_\alpha) +
\Lambda_\beta\Theta(\rho-\rho_\beta) + \Lambda_\gamma\Theta(\rho-\rho_\gamma)
\big],
\end{eqnarray}
where $\Lambda_i$ is the residue of $I_{\alpha,\beta,\gamma}(\lambda)$ at $
\lambda_i$ with $i=\alpha, \beta, \gamma$, and $\Theta$ denotes the usual
step function.

\begin{itemize}
\item  The cases $\gamma =\alpha $ or $\gamma =\beta $
\end{itemize}

\begin{eqnarray}
&&\int d \lambda I_{\alpha,\beta,\alpha}(\lambda) = \int |\lambda|dz 
\frac{1}{(|\lambda|z -\lambda_{\alpha})^2 (|\lambda|z -\lambda_{\beta})}
  \nonumber\\
&& \quad \qquad \qquad \ \ \ = {\frac{1}{{\lambda_{\alpha\beta}^2}}}
[-\Theta(\rho-\rho_\alpha) +\Theta(\rho-\rho_\beta)],
\end{eqnarray}
where $\lambda_{\alpha\beta}=\lambda_{\alpha}-\lambda_\beta$.

Remembering the ordering of $\rho _{\alpha }$ we can now write the
superpotential as 
\begin{eqnarray}
&&W=-2\pi iR\sum_{\alpha <\beta }(c_{\alpha }s_{\beta }-c_{\beta }s_{\alpha
})\int dg_{\alpha }\wedge dg_{\beta }\bigg[ {\frac{(a_{\alpha }-a_{\beta })}{
\lambda _{\alpha \beta }^{2}}}\Theta _{\alpha \beta }(\rho )  \nonumber \\
&&\ \ +\sum_{\gamma \ne \alpha ,\beta }^{n}a_{\gamma }\left( \Lambda
_{\alpha }\Theta (\rho -\rho _{\alpha })+\Lambda _{\beta }\Theta (\rho -\rho
_{\beta })+\Lambda _{\gamma }\Theta (\rho -\rho _{\gamma })\right) \bigg],
\end{eqnarray}
where $\Theta _{\alpha \beta }(\rho )$ is the step function that is 1
between $\rho _{\alpha }$ and $\rho _{\beta }$ and 0 otherwise. We can also
evaluate the rest of the integral. From the construction of the homotopy
functions we notice following fact, 
\[
\int_{C_{\gamma }}g_{\alpha }dg_{\beta }=\delta _{\alpha \gamma }, 
\]
where $C_{\gamma }$ is a line along the $\rho =\rho _{\gamma }$ as $\sigma $
varies from 0 to 1. By this and the Stokes' theorem it is easy to evaluate
the integral 
\[
\int_{\gamma \delta }dg_{\alpha }\wedge dg_{\beta }={\frac{1}{2}}(\delta
_{\alpha \gamma }+\delta _{\beta \delta }-\delta _{\alpha \delta }-\delta
_{\beta \gamma }), 
\]
where the integral is over a band of the strip defined by $\rho _{\delta
}<\rho <\rho _{\gamma },0\le \sigma \le 1$. Thus, the final result is 
\[
W=-2\pi iR\sum_{\alpha <\beta }(c_{\alpha }s_{\beta }-c_{\beta }s_{\alpha
})\left[ {\frac{(a_{\alpha }-a_{\beta })}{\lambda _{\alpha \beta }^{2}}}+{
\frac{1}{2}}{\frac{1}{\lambda _{\alpha \beta }}}\sum_{\gamma \neq \alpha
,\beta }a_{\gamma }\left( {\frac{1}{\lambda _{\gamma \alpha }}}+{\frac{1}
{\lambda _{\gamma \beta }}}\right) \right] , 
\]
where $\alpha ,\beta ,\gamma $ run from 0 to $n$.

If we choose the parametrization where first NS fivebrane is identified with
the large $\lambda$ region, it is given by 
\begin{eqnarray}
&&v=\lambda+\sum_{\alpha=1}^{n} \frac{c_{\alpha}}{\lambda -\lambda_{\alpha}},
\nonumber \\
&&w=\sum_{\alpha=1}^{n} \frac{s_{\alpha}}{\lambda -\lambda_{\alpha}}, 
\nonumber \\
&&t=\prod_{\alpha=1}^{n}(\lambda -\lambda_{\alpha})^{-a_{\alpha}}.
\end{eqnarray}
Notice that the $\alpha=0^{{\rm th}}$ components in $w$ and $t$ are
truncated. Sometimes, this parametrization makes the discussion of the
physics more intuitive. For example, when we discuss the rotations of the
branes, one of the branes must be fixed and this distinguished one is
located at $\lambda=\infty$. Almost the same calculation gives us the
superpotential, 
\begin{eqnarray}
&&W= -2\pi i R \sum_{\alpha <\beta}^n (c_\alpha s_\beta-c_\beta s_\alpha)
\left[ {\frac{(a_\alpha-a_\beta)}{\lambda_{\alpha\beta}^2}}+ {\frac{1}{2}} 
{\frac{1}{\lambda_{\alpha\beta}}} \sum_{\gamma\neq\alpha,\beta}a_\gamma
\left({\frac{1}{\lambda_{\gamma\alpha}}} +{\frac{1}{\lambda_{\gamma\beta} }}
\right)\right]  \nonumber \\
&& \ \ \ \ \ \ -2\pi iR \left[ (\sum_{\gamma=1}^{n} a_\gamma)
(\sum_{\beta=1}^n s_\beta)- {\frac{1}{2}}\sum_{\gamma\ne\beta}{\frac{
a_\gamma \lambda_\gamma s_\beta }{\lambda_{\gamma\beta} }} \right].
\end{eqnarray}
Here $\alpha,\beta,\gamma$ run from 1 to $n$.

The above discussion can easily be generalized to the cases with $N_f$
hypermultiplets, corresponding to $N_f$ D sixbranes or semi-infinite D
fourbranes. To illustrate this let us consider a situation where the $N_f$
of semi-infinite D fourbranes are attached to the right hand side of the
second NS fivebrane. The curve for such a configuration had already been
written down\cite{hoo}. In the parametric form, it is
given by 
\begin{eqnarray}
&&v={\frac{(\lambda-\lambda_+)(\lambda-\lambda_-)}{\mu \lambda}},  \nonumber
\\
&&w=\lambda,  \nonumber \\
&&t=\mu^{-N_c}\lambda^{N_c-N_f} (\lambda-\lambda_+)^r
(\lambda-\lambda_-)^{N_f-r},
\end{eqnarray}
where $\lambda_\pm$ are the two solutions of $v=0$, and $\mu$ is bare mass
for the adjoint chiral multiplet. Employing the same method as above, we get
following superpotential: 
\begin{eqnarray}
&&W=2\pi i {\frac{R\lambda_+\lambda_- }{\mu}} [2(N_c-N_f)+(N_f-r) +r], 
\nonumber \\
&& \ \ \ \ = 2\pi i {\frac{R\lambda_+\lambda_- }{\mu}}(2N_c-N_f).
\end{eqnarray}
This precisely agrees with the minimum value of $N=1$ gauge theory
superpotential ($N_f < N_c$) given as follows\cite{ads,hoo}; 
\begin{equation}
W_{{\rm {eff}}}= (N_{c}-N_{f}) \left(\frac{\Lambda _{N=1}^{3N_c-N_f}}{\det M 
}\right)^{1/(N_c-N_f)} +\frac{1}{2\mu } \left( {\rm Tr}(M^2)-\frac{1}{N_c}(
{\rm Tr}M)^2\right),
\end{equation}
where $\Lambda _{N=1}$ is the dynamical scale of $N=1$ SQCD. More
explicitly, 
\begin{eqnarray}
W_{{\rm Gauge-theory}}=-{\frac{1}{4\pi i}}W_{{\rm M-theory}},
\end{eqnarray}
with the identification $\lambda_\pm= m_\pm$, where $m_\pm$ are the only two
possibly different eigenvalues of the meson matrix $M$ whose elements are ${
\tilde Q}_i^a Q_b^i$. This is the first concrete evidence that Witten's
proposal for the superpotential is correct. There is an explicit 
rotation angle dependence through $\mu\sim \tan\theta$. 
The dependence on $r$ comes only through $\lambda_\pm$.

\section{Discussion}
In this paper we found a parametric representation of the brane
configurations corresponding to the $N=1$ SQCD. Extending the Witten's
method, we developed a general formalism to calculate the superpotentials
and compared with known gauge theory results.

The comparison with gauge theory result might be interesting when we
consider gauge theories for product groups. As mentioned earlier, one
motivation for this comes from the study of configurations of $k$
`coincident' NS fivebranes linked to an `coincident' NS fivebrane linked by $
N_c$ D fourbranes, and have the form of the superpotential for that
configuration\cite{egkrs}. The gauge theory has $SU(N_c)$ gauge symmetry.
Now when the NS fivebranes separate it corresponds to, in the field theory,
resolution of the singularity in the superpotential of the form $W=X^{k+1} + 
$lower powers\cite{kutasov}. Then the gauge symmetry is spontaneously broken
to $SU(N_c) \rightarrow \prod SU(r_i)$, where $\sum_i r_i = N_c$. Of course
the detailed form of the potential depends on the deformation parameters
away from the coincident limit.

Also more thorough understanding is necessary on the role of the complex
volume form which appears in the definition of the superpotential. The
implication of the superpotential in relation to the symmetry enhancement
when the branes collapse is not fully understood. These are currently being
investigated.

\vskip 2cm \leftline{\bf Acknowledgement} \noindent KhO would like to thank
POSTECH, SJS and SN to APCTP for hospitality. SJS also likes to thank APCTP
for financial support and PIMS for the hospitality during his stay at UBC
where part of the work is done. This work is supported in part by Ministry
of Education(BSRI-96-2441 for SJS, BSRI-97-2442 for SN), by KOSEF
(971-0201-001-2 for SJS, 961-0201-001-2 for SN), and by CTP/SNU through the
SRC program of KOSEF(SN). Finally, we would like to thank S. Hyun for his
participation in the early stage.

\end{document}